\documentclass[pra,twocolumn,showpacs,preprintnumbers,amsmath,amssymb]{revtex4-1}
\makeatother
\usepackage[dvips]{graphicx}
\usepackage{amsmath}
\usepackage{amsbsy}
\usepackage{color}
\DeclareGraphicsExtensions{.pdf,.eps,.png,.jpg,.mps}

\definecolor{textcolor}{cmyk}{0,0,0,1}
\definecolor{magenta}{rgb}{1,0,1}
\definecolor{green}{rgb}{0,1,0}
\definecolor{red}{rgb}{1,0,0}

\begin{document}

\title{
Divacancy-induced Ferromagnetism in Graphene Nanoribbons }
\author{W. Jask\'olski}
\affiliation{
Institute of Physics, Faculty of Physics, Astronomy and Informatics, Nicolaus Copernicus University, Grudziadzka 5, 87-100 Torun, Poland}
 \author{Leonor Chico}
\affiliation{Departamento de Teor\'{\i}a y Simulaci\'on de Materiales, Instituto de Ciencia de Materiales de Madrid (ICMM), Consejo Superior de Investigaciones Cient\'{\i}ficas (CSIC), C/ Sor Juana In\'es de la Cruz 3,
28049 Madrid, Spain}
\author{A. Ayuela}
\affiliation{Centro de F\'{\i}sica de Materiales CFM-MPC CSIC-UPV/EHU, Donostia 
International Physics Center (DIPC), Departamento de F\'{\i}sica de Materiales, Fac. de Qu\'{\i}micas, UPV-EHU, 20018 San Sebasti\'an, Spain
}

\date{\today}

\begin{abstract}
Zigzag graphene nanoribbons have spin-polarized edges, anti-ferromagnetically coupled in the ground state with total spin zero. 
Customarily, these ribbons are made ferromagnetic by producing an imbalance between the two sublattices. 
Here we show that zigzag ribbons can be ferromagnetic due to the presence of reconstructed divacancies near 
one edge. This effect takes place despite 
the divacancies are produced by removing two atoms from opposite sublattices,
 being balanced before reconstruction to 5-8-5 defects. 
 We demonstrate that there is a strong interaction between the 
 defect-localized and 
  edge 
  bands 
  which 
  mix and split away from the Fermi level. This splitting is asymmetric,  yielding a net edge spin-polarization. 
  Therefore, the
  formation of reconstructed divacancies close to the edges of the nanoribbons can be a practical 
  way to make them 
  partially 
  ferromagnetic.
\end{abstract}

\pacs{73.22.-f, 73.63.-b}

\maketitle

\section{Introduction} 
Magnetism in zigzag graphene nanoribbons (ZGNR) 
is related to  
edge-localized states, 
which appear as 
 two flat bands at the Fermi energy ($E_F$) in a simple noninteracting model. 
  In fact, the electron interaction splits these bands, so the edges are antiferromagnetically coupled with total spin zero \cite{son_nature,son_prl2006}. 
 This magnetic behavior is rather general, 
  because similar localized bands are also present in any non-armchair graphene ribbon \cite{fujita,nakada,our_prb2011}. When the edges of the nanoribbon are identical, all the bands remain spin-degenerate. For dissimilar edges with sublattice balance, the spin splitting may be different for each edge \cite{our_ssc_2014}, but the ribbons have total spin zero.
In order to exploit 
spin effects in ZGNRs 
for applications, spin degeneracy  
should be lifted, so 
uncompensated spin channels are obtained. 
Such splitting can be achieved under a
  strong external electric field  \cite{son_nature,mananes} or by chemical attack \cite{hod}.

In general, one way to attain  ferromagnetic 
graphene nanostructures is to impose 
 sublattice imbalance. According to Lieb's theorem, a bipartite lattice has a total spin moment proportional to the
 difference of the number of atoms belonging to the two sublattices \cite{lieb}. 
For instance, ZGNRs with one decorated edge of Klein-type 
atoms \cite{klein,book_CBM},  triangular graphene nanoislands 
  \cite{jfr_palacios_prl2007}, and graphene systems with vacancies that remove a different number of nodes from each sublattice \cite{palacios_jfr_prb2008,palacios_jfr_bry_fertig_sst2010,cnr,lps} 
have a non-zero spin due to the imbalance.  
In this work we 
show another way of producing a net magnetic moment in 
zigzag graphene nanoribbons  by including reconstructed divacancies.

We consider divacancies produced by the removal of two neighbor carbon atoms, so that the two sublattices
 are balanced.
 They rebuild into the so-called 5-8-5 defects, composed of an octagon and a pair of pentagons 
  which mix the two sublattices. 
Divacancies may naturally appear as stable defects during growth or can be created on purpose by electron or ion irradiation 
\cite{hashimoto_nature2004,kotakoski_prl2011,kim_prb2011,robertson_nc2012,ugeda_prl2012,ugeda2}. 
They are the source of defect-localized states with energies close to $E_F$, as it was  recently shown for the case of 
semiconducting 
armchair ribbons \cite{our_app2013}.
%
%
Since divacancies do not introduce sublattice imbalance, they have not been regarded to this date
as possible sources of magnetization in graphene. 
However, we show here that when these defects are present in zigzag nanoribbons, 
they give rise to localized states which may interact with those originated from the zigzag edges, 
so they can lead   
to spin effects and ribbon magnetization. 

Two previous 
calculations for 5-8-5 defects in 
ZGNRs presented results in apparent contradiction, showing either 
zero spin polarization \cite{topsakal_prb2008},
or spin-polarized transport in ribbons with narrow widths \cite{oeiras_prb2009}. 
The issue of 
whether these defective nanoribbons 
are ferromagnetic or not was not addressed in those works. In principle, one could 
interpret that spin polarization arose in narrow ribbons because of size effects. 

In order to clarify this point, 
in this work we perform a systematic study of 
 the magnetic behavior of ZGNRs with reconstructed divacancies. 
 We have found that, although these divacancies arise from lattice-balanced defects, can nevertheless produce a net magnetic moment in zigzag nanoribbons. 
This happens when they are located close to the zigzag ribbon edge. We attribute the appearance of a nonzero spin to the strong interaction between edge and divacancy  states. 
%
%

   \begin{figure}[thpb]
      \centering
\includegraphics[width=5.5cm]{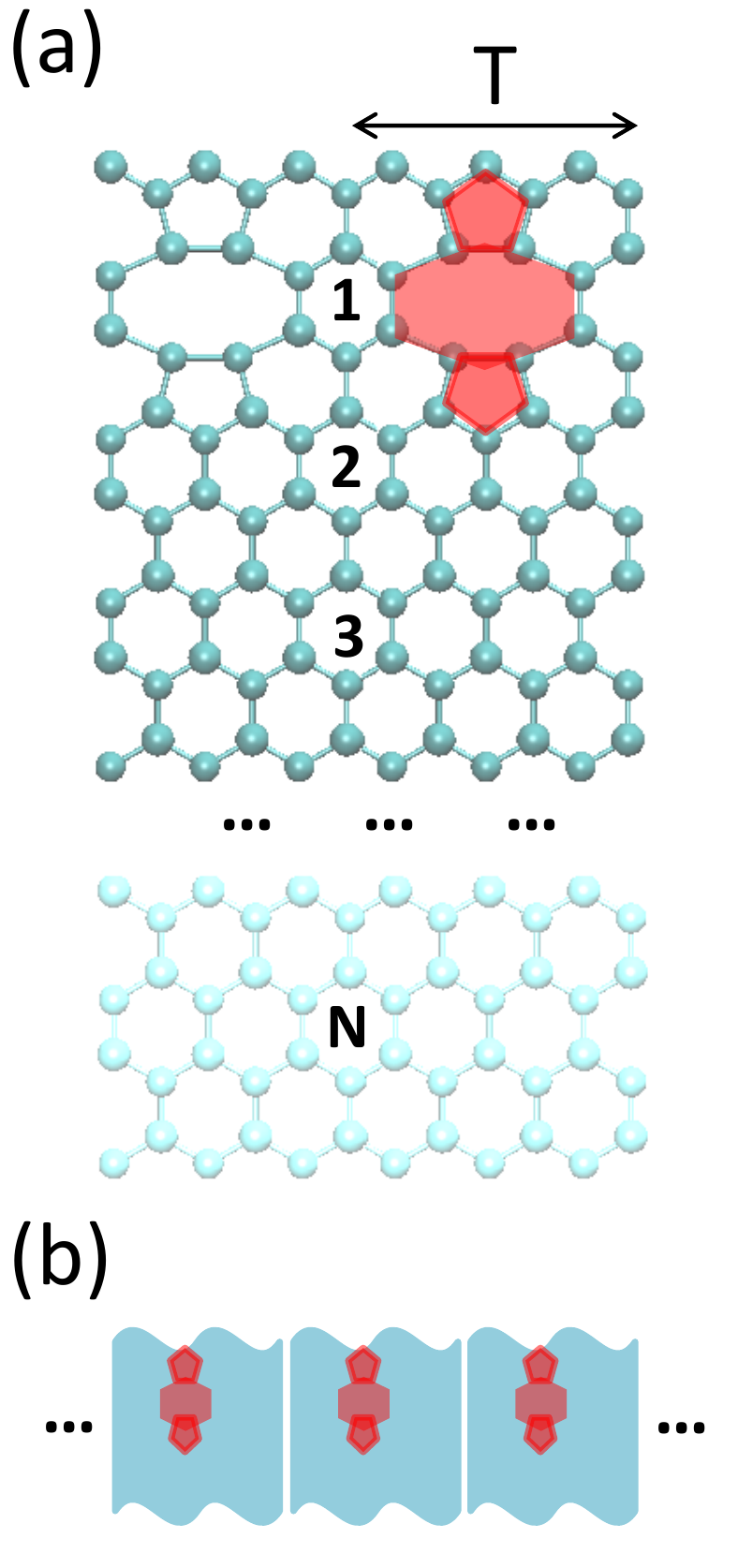}
\caption{\label{fig:fig1} 
(Color online) 
(a) Divacancies in ZGNRs. The position of the defect 
in the ribbon is given by 
the integer 
$N$. (b) Schematic 
drawing of the 
 periodically
 placed defects along the ribbon
 forming a superlattice. 
The translation period  
$T$ 
spans the length 
of the unit cell. } 
   \end{figure}

 We have examined systematically how the magnetic properties of ZGNR depend on the position of 5-8-5 defects. We show that when defects are
centrally located in wide ZGNRs, the ribbons 
 have zero net magnetic moment.  However, when they are placed close to one of the zigzag edges, the defect-localized 
 and the nearby edge bands interact,  
 so they mix and split in energy. 
The zero energy band 
corresponding to
the other edge situated farther from the defect remains unmixed. 
The inclusion of 
electron-electron interaction results in the spin splitting of all these bands. 
 The aforementioned unmixed band 
  is symmetrically split around $E_F$, while the 
  hybridized 
  defect-edge bands are asymmetrically
  split, yielding 
  a  non-zero 
  net 
  magnetization. 
We propose that
 the 
 production 
 of reconstructed divacancies 
 by techniques such as ion bombardment  may produce
 magnetic ribbons. 
  As 
  the 
  one of the most abundant defects in ZGNRs 
  are divacancies \cite{wang}, 
  which are 
  preferentially found 
   at the edges \cite{oeiras_prb2009}, this defect engineering could be a feasible way to produce spin-polarized ZGNRs.

\section{Model and systems studied} 
We study 
reconstructed 
divacancies in wide zigzag graphene nanoribbons. The ribbon width $W$ is defined \cite{nakada_fujita} as the number of carbon dimers across the ZGNR. 
The  divacancies
 are located 
at different positions 
$N$, measured in units of two carbon 
dimers from the edge of the ribbon, 
as shown in Fig.  \ref{fig:fig1} (a). The 5-8-5 defects are 
periodically situated 
 in an infinite ZGNR,
 as 
schematically depicted in Fig. \ref{fig:fig1} (b). 
The translation period $T$ of the ribbon is defined as the number of zigzag edge nodes in the unit cell. The electronic properties are 
calculated with a one $\pi$-orbital tight-binding (TB) model. 
The electron-electron 
interaction is 
considered 
within a Hubbard model solved in the mean-field approximation.
We choose this approach in order to calculate large 
 unit cells, which are not feasible with first-principles methods.  
 We assume all hoppings $t$ to be equal throughout the ribbon. 
We previously tested this approach for the study of the magnetic properties in graphene with topological defects using the hopping parameter $t=-2.7$ eV and the Coulomb interaction term $U=3$ eV \cite{our_prb2013,our_ssc_2014}.


\section{
Results} 

\begin{figure}[thpb]
      \centering
\includegraphics[width=8.5cm]{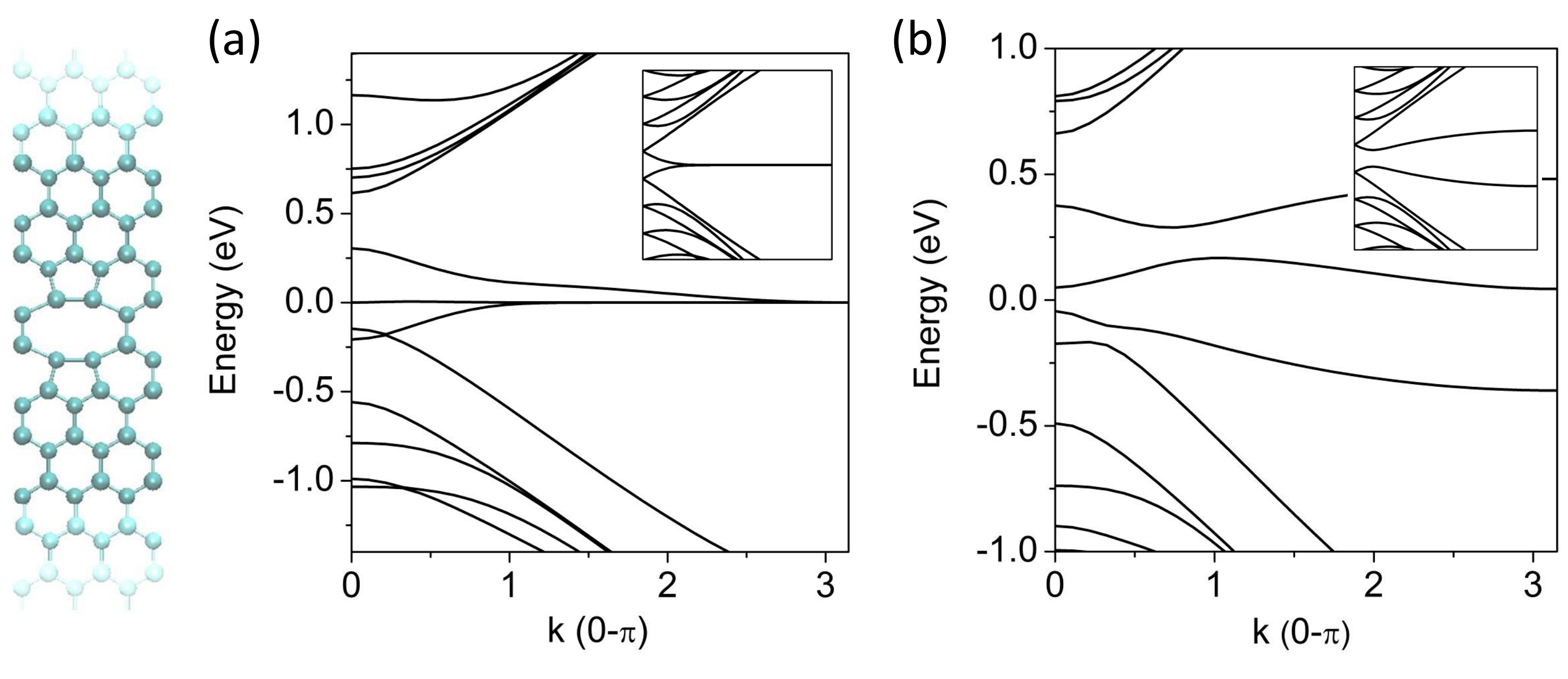}
\caption{\label{fig:fig2} (Color online) 
Bands of a ZGNR with $W=19$ and $T=3$ with 5-8-5 defects located at the center of the ribbon, calculated in (a) the TB approximation and (b) the Hubbard model. For comparison, the corresponding spectra of pure ZGNR folded three times ($T=3$) are included as insets. The Fermi level lies at $E=0$.  Notice that in (b) the zero energy edge bands are spin-split, but they remain spin degenerate with no spin polarization; the defect band is spin-degenerate. } 
   \end{figure}

We first consider centrally located 5-8-5 defects,  
periodically placed along a wide ribbon with $W=19$ and $T=3$. It is the smallest periodicity of a ZGNR with horizontally placed defects separated by at least one hexagon. 
Note that for $T=2$ the 
consecutive 
octagon-pentagon pair defects
 form a defect line, which was studied elsewhere \cite{our_prb2013,Song2012,Jiang2012,Kan2012,Hu2012}. 
The energy spectra calculated within the  TB approximation and the Hubbard model are presented in
  Figs. \ref{fig:fig2}(a) and (b), respectively.  The spectra are not symmetric with respect to 
   $E=0$ because of the electron-hole symmetry breaking induced by
   the pentagons. The insets show the band structure of a ZGNR with $T=3$, i.e., the three times folded spectrum of the pure (1,0) ZGNR with  
the Dirac point at $k=0$. It has a pair of zero-energy bands extending in the entire zone 
in the TB approximation (Fig. \ref{fig:fig2} (a)), 
which are split when the electron-electron interaction is included
 (Fig. \ref{fig:fig2} (b)) 
 \cite{son_prl2006}.
The 5-8-5 defects introduce divacancy-localized states, which in the TB approximation form a flat band exactly at $E=0$, as shown in Fig. \ref{fig:fig2} (a) \cite{our_app2013}. When the Coulomb interaction is considered, as in Fig. \ref{fig:fig2} (b), 
 the two edge localized bands are spin-split in the same way 
 as in the pristine ZGNR. 
 The unoccupied defect-localized band remains spin-degenerate. As the defect is symmetric about the
  center of the ribbon,  the ground state remains anti-ferromagnetic.

\begin{figure}[thpb]
      \centering
\includegraphics[width=9cm]{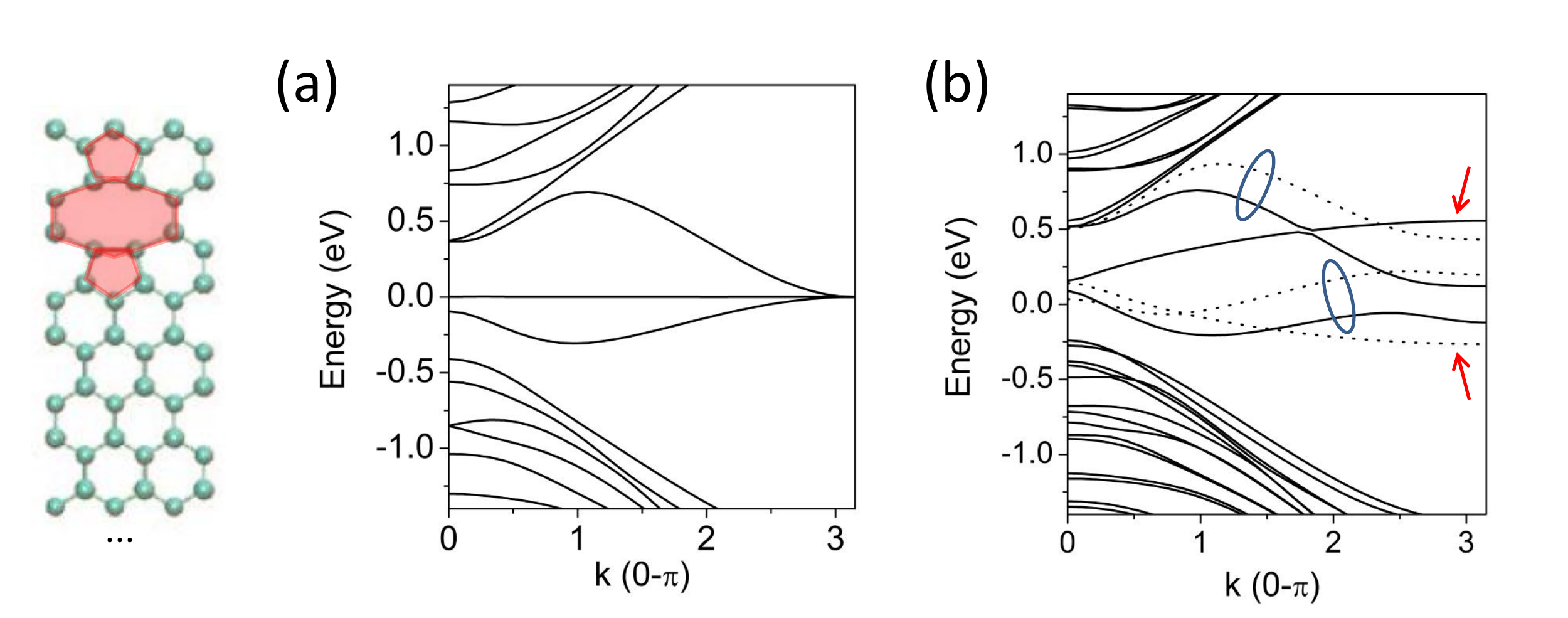}
\caption{\label{fig:fig3} 
(Color online) Bands of a ZGNR with $W=19$ and 5-8-5 defects placed close to the upper edge ($N=1$) and separated by the translation vector $T=3$, calculated in the TB (a) and Hubbard (b) models. Spin-down bands are denoted by dotted lines, spin-up by solid lines. Arrows 
mark 
the up and down spin bands localized at the lower edge. 
Ellipses mark the spin-split bands of the bonding and anti-bonding combinations between the upper edge and the defect-localized states. 
 }
   \end{figure}

More interesting is when we move the 5-8-5 defect close ($N=1$) to one of the edges of the ribbon, e.g., 
 the upper one. The energy spectra calculated in the TB and Hubbard models are shown in Figs. \ref{fig:fig3} (a) and (b), respectively. In the TB approximation the flat band at $E=0$ is localized at the lower edge, and it remains unaffected by divacancies. However, the states localized at the upper edge 
strongly interact with the defect-localized states; they 
hybridize and split. The bonding combination of these states is the band 
below $E=0$, while the anti-bonding combination is 
unoccupied. All the bands are spin-degenerate. 
The inclusion of 
electron-electron interaction
lifts the spin degeneracy and 
significantly modifies the spectrum, as it can be seen in Fig. \ref{fig:fig3} (b). 
The bands localized at the lower edge
are marked 
in Fig. \ref{fig:fig3} (b) by arrows. The spin-down polarized states are almost fully occupied.
The 
bonding 
and anti-bonding combinations between the upper-edge and defect-localized 
bands 
also spin-split. However, their 
splitting 
is weaker because of the defect-edge mixing. The spin-split pairs are marked with ellipses. 
In the TB approximation the bonding combination 
is situated closer to 
$E_F$ than its anti-bonding counterpart. Now, when the 
Hubbard term 
 is included, 
 spin-up and spin-down bands cross the Fermi level at different $k$-values. This produces a non-zero 
  final spin polarization, 
  about $ 0.2$ $\mu_B$. Note that this result does not contradict Lieb's theorem: although the lattice was balanced before
  reconstruction, the mixing of sublattices produced by the topological defects makes the theorem inapplicable to this case. 
  Significantly, ZGNRs on one hand and 5-8-5 defects on the other hand have a total spin zero. However, 
  when the defect is placed close to one of the edges, a net spin appears due to the asymmetrical band splitting produced by
  the defect-edge interaction.

\begin{figure}[thpb]
      \centering
\includegraphics[width=8.5cm]{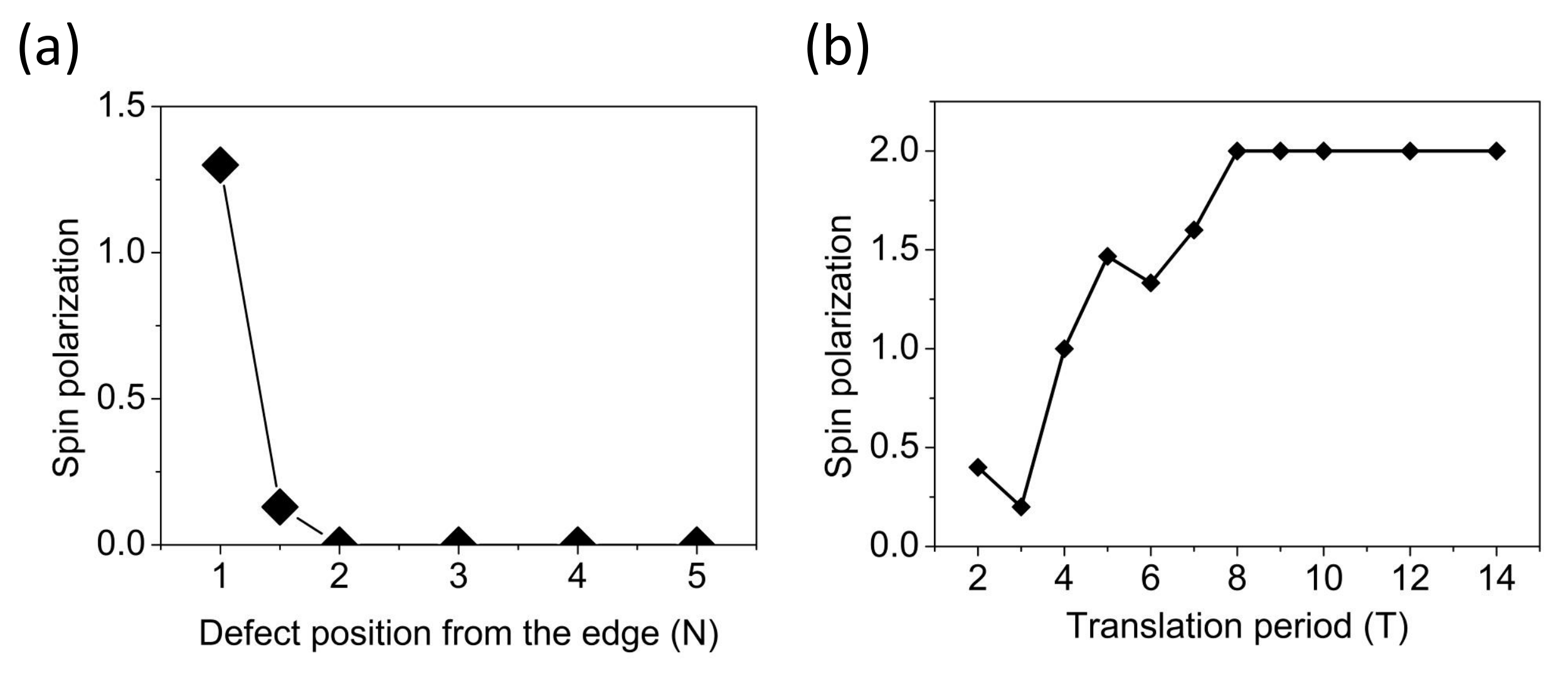}
\caption{\label{fig:fig4} 
Dependence of spin polarization on (a) the position $N$ of defects 
with respect to the edge of 
the ribbon,
and (b) the translation period $T$, for a wide ribbon of $W=39$. 
}
   \end{figure}

Figure \ref{fig:fig4} (a) shows how the spin polarization depends on the position of the defect 
with respect to the edge of 
the ribbon. 
 We 
 consider a 
 ribbon with $T=6$ and $W=39$, which is wide enough to have the defect in several sites between the center of the ribbon 
 and its edge. 
 When the defect is situated close to the edge, i.e. $N=1$, the spin polarization is $ 1.3$ $\mu_B$.  Moving the defect towards the 
 center makes the polarization decrease rapidly to zero. 
This may explain why no magnetization was reported in the study
 presented 
  in Ref. \onlinecite{topsakal_prb2008} for 
ZGNR with slightly off-center divacancies. Another 
work  \cite{oeiras_prb2009} gives an example of a very narrow ribbon, not large enough to
distinguish 
 the magnetic polarization effect induced by such defects from 
 that caused by the edges themselves.
We have 
systematically studied 
how the spin polarization depends on the translation period $T$ for the ribbon of width $W=39$, as shown in Fig. \ref{fig:fig4} (b). When $T$ increases, the ribbon polarization 
also
increases (albeit non monotonically), and saturates 
for large $T$ 
  at $2$ $\mu_B$. 
  

In order to understand the values of spin polarization presented above, we have also studied in more detail the energy spectra of ZGNR with larger $T$. When the defects are situated at the center of the ribbon, no
  spin polarization is observed 
  for any $T$.
  In order to
  compare the results with those 
  with $T=3$ presented above, in which the Dirac point is at $k=0$, we 
  choose 
   $T$ 
   to be a
    multiple of three. The smallest translation period
     for which the polarization converges to
      $2$ $\mu_B$
      is $T=9$. 
The TB energy spectrum of
a
 pure ZGNR (9,0) 
 has 6 flat bands at $E=0$ \cite{our_prb2011}. Three of them are localized at the lower edge, and another trio is 
 at the upper edge. 

   \begin{figure*}[thpb]
      \centering
\includegraphics[width=14cm]{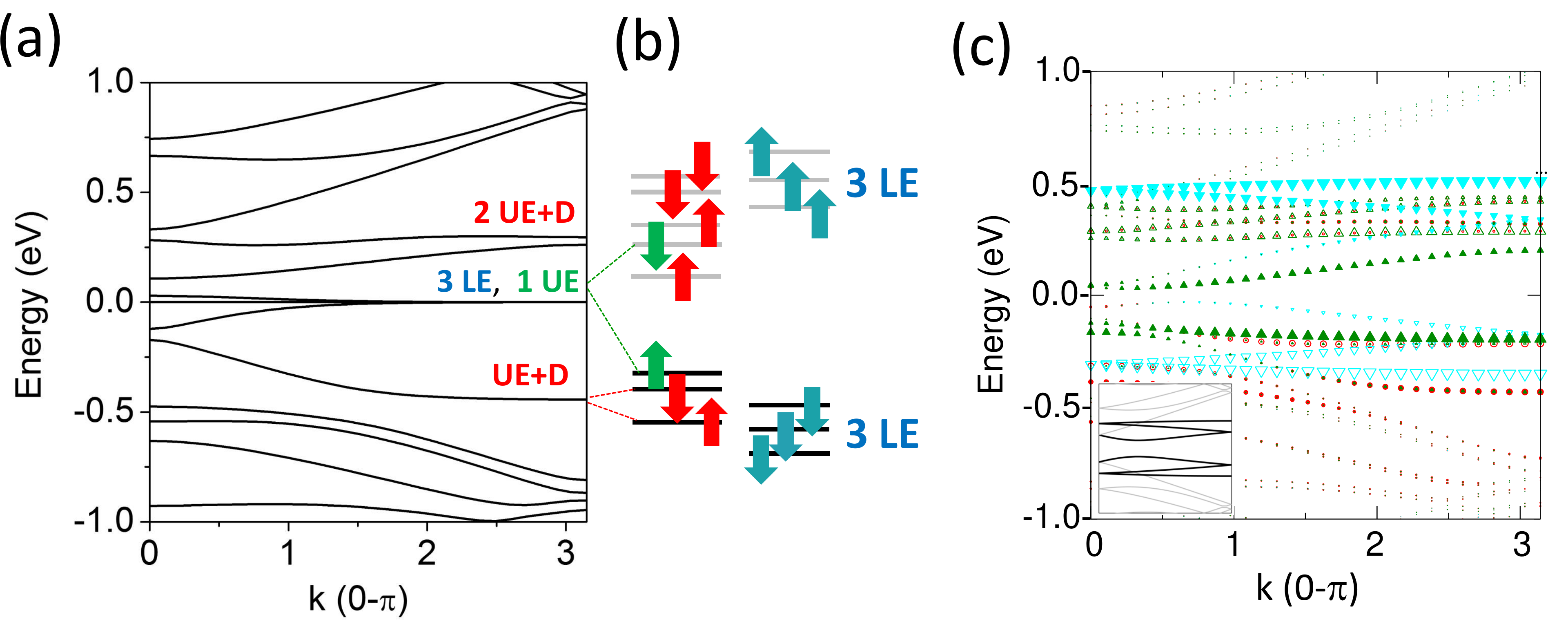}
\caption{\label{fig:fig5} 
(Color online) (a) Bands of a ZGNR of $W=19$, $T=9$ and a 5-8-5 defect placed close to the edge ($N=1$), calculated with the TB model; all the bands are spin-degenerate. Bands localized almost exclusively at the upper and lower edge nodes for $k=\pi$ are denoted by UE (green) and LE (blue),  respectively. 
 Bands localized both at the upper edge and at the defect are marked as UE+D (red). 
 (b) 
  Schematic diagram showing how 
 the bands are spin-split when the Coulomb interaction is considered. (c) Band structure calculated with the Hubbard model. Spin-up (full 
 symbols) and spin-down (empty 
 symbols) bands at the nodes of the upper edge (up triangles, green),  the lower edge (down triangles, blue) and the defect atoms (circles, red). 
The 
symbol sizes are
proportional to the probability density 
at the defect atoms, upper edge atoms and lower edge atoms. Inset:  Hubbard bands of the pure ZGNR (9,0).
}
  \end{figure*}

When the 5-8-5 defects 
are close to the upper edge, 
defect and upper-edge bands interact, 
so they mix and 
 split. Fig. \ref{fig:fig5} shows the TB energy bands for the $W=19$ and $T=9$ ZGNRs 
 with a 5-8-5 defect 
 close to the 
 edge. 
There are four flat bands near to the Fermi level. Three bands  
are localized at the lower 
edge (LE) and 
one 
at the upper edge (UE). 
 These states are 
 mostly 
  localized at the edge nodes and have 
a negligible overlap with the defect atoms. The two remaining upper edge 
bands
hybridize with the defect
band, yielding 
 three bands (UE+D) of mixed upper edge-defect character,  with only one occupied and all away from $E_F$. 

A 
diagram 
showing how these bands spin-split when the Coulomb interaction is taken into account is presented in Fig. \ref{fig:fig5} (b).
The LE bands (blue) are strongly split so they are
 fully spin-polarized; we take the occupied spin as the down projection. 
  The 
 unperturbed
 $E=0$ UE band (green) is 
 split with a spin opposite to the 
  LE bands, as expected. The UE+D bands are spin-split more weakly because of the sublattice mixing at the defect. Consequently, we have four spin-down and two spin-up occupied states, summing up to $2$ $\mu_B$. Calculations employing the Hubbard model confirm this
 picture, 
 as displayed by the energy bands in 
 Fig. \ref{fig:fig5} (c).  
  We have checked that 
our results are 
  robust, i.e., 
 independent on the Coulomb term 
 for a 
 wide range of $U$ values. 

   \begin{figure}[thpb]
      \centering
\includegraphics[width=8.5cm]{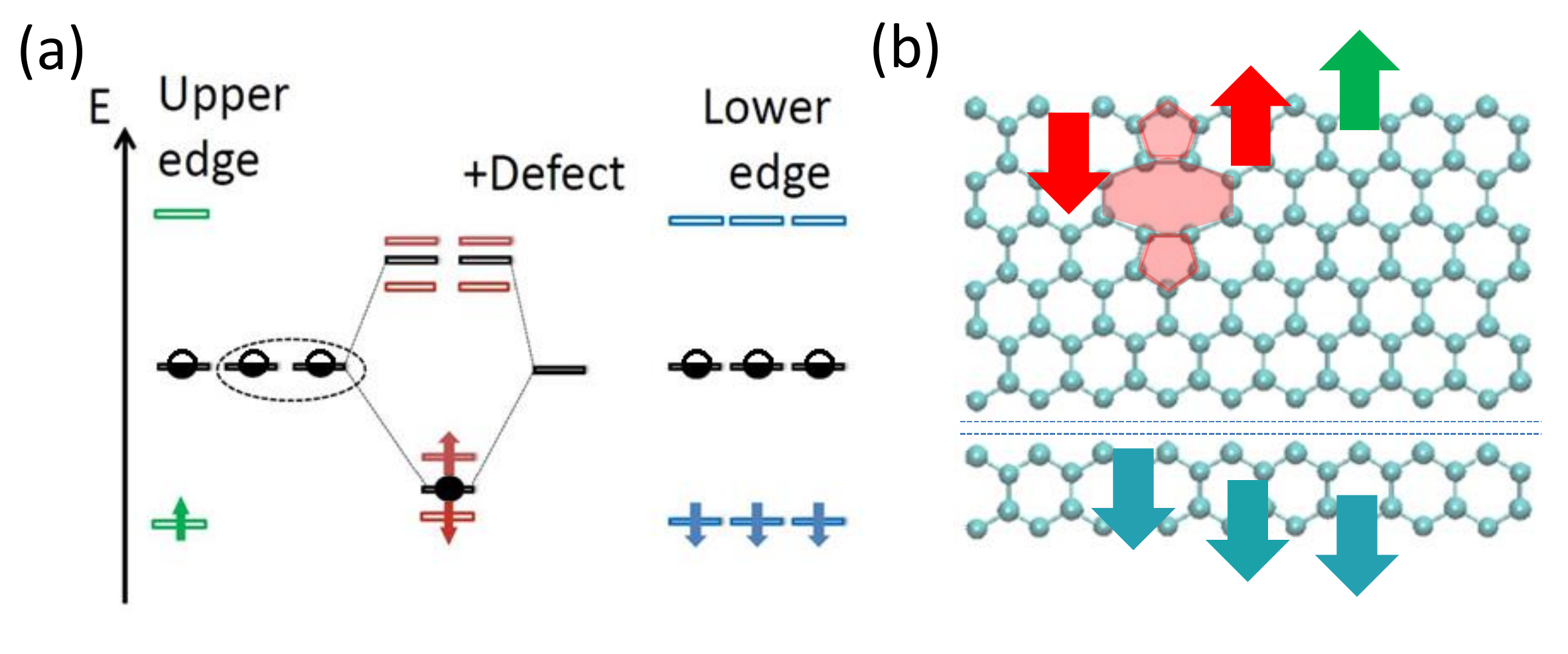}
\caption{\label{fig:fig6} 
(Color online) %
(a) Schematic energy diagram showing how the defect-localized state mixes with two  
 upper-edge localized states. 
The TB levels are shown in black; filled and half-filled dots indicate that the state is fully or half-occupied, respectively. 
Electron interaction splits these levels; colors indicate their localization as in Fig. 5 (b).
  Occupied states are represented with an arrow indicating the spin direction. (b) 
  Diagram illustrating the final distribution of occupied spin states 
  due to divacancy placed close to the edge.
}
   \end{figure}

The energy and spatial schemes presented in Fig. \ref{fig:fig6}  describe these edge-defect interactions in detail. The edge bands of the pure ZGNR (1,0), which extend from $k=2\pi/3$ to  $k=\pi$, contribute in average with one electron for 
every
three edge nodes. For $T=9$ 
this 
band folds into three edge bands.
For the divacancy close to the upper edge, the defect 
band 
hybridizes with two edge
bands, giving 
 three 
 bands 
 away from $E_F$, as shown in Fig. \ref{fig:fig6} (a). 
 The defect does not 
 mix with the remaining upper edge band 
 (green), 
 because it 
 stems 
 from the 
 states close to $k=\pi$ 
 of the unfolded edge band of the ZGNR (1,0): as it is
mostly localized at the edge atoms, 
it has a small overlap with the
 defect. 
 Likewise, the lower edge discrete states (blue) are also unaffected by the 
divacancy 
 due to the spatial separation. These unchanged states are spin-split as for a pure ZGNR. However, the spin splitting of the hybridized states (red) is much weaker, with a state below $E_F$ occupied 
for
both spin polarizations. 
For even larger $T$, 
 the spins of the extra occupied upper edge 
 states 
 far from the defect cancel with the spins of the occupied lower edge states. Thus, an isolated straight divacancy in an infinite ZGNR has a total spin polarization equal to $ 2$ $\mu_B$. We have also checked that tilted vacancies have a similar behavior, although the total magnetic moments tend to be reduced.

This divacancy-induced magnetism at the edge of zigzag ribbons is now brought into contact with experiments. We propose that  ion irradiation 
of zigzag ribbons could be employed to create divacancies. This technique is used nowadays to produce vacancies in graphene \cite{ugeda_prl2012}. 
In fact, vacancies 
are mobile and cluster in the form of divacancies \cite{ugeda2}. Note that for nanoribbons, it is more energetically favorable for these
 %
 vacancies 
 to move 
 close to the edge because of its lower coordination, where they can coalesce in the more stable and abundant divacancies \cite{wang}.

\section{Conclusions} 
%
%
In summary, we have studied the electronic and magnetic properties of ZGNRs with reconstructed divacancies, which can be viewed as the removal of two neighbor carbon atoms from different sublattices before reconstruction to 5-8-5 defects.  Although 5-8-5 defects stem from lattice-balanced vacancies, they can give rise to a net spin magnetic moment. We have shown that a nonzero magnetization arises when the 
defect is located close to the edges of the zigzag ribbon.
%
%
The 5-8-5 defects introduce localized states with energies close to $E_F$. 
When they are located at the center of the ribbon, the total spin polarization is zero, keeping the magnetic edge configuration of pristine ribbons. 
However, when the defects are placed closer to one of the edges of the ribbon, 
the defect band  
 interacts with the edge-localized band,  
 so they hybridize 
 and split asymmetrically from $E_F$. 
 States localized at the other edge remain strongly spin-split, leading to a net spin polarization and
 spontaneous magnetization of the ribbon, despite they are derived from systems with  
 balanced sublattices before reconstruction. The total magnetic moment saturates for large periods to 
 a value of 2 $\mu_B$. 
%
%
Finally, 
we have 
also
%
%
clarified the apparent contradiction between previous works, namely, the absence of spin polarization shown for some defective ribbons, in contrast with the 
obtention of spin-polarized currents in similar systems \cite{oeiras_prb2009,topsakal_prb2008}. In narrow ribbons, divacancies are naturally close to edges, so 
a spin-polarized current may arise. In wider ribbons, divacancies situated at the central region of the ribbon do not produce such spin polarization. 

 Our findings indicate that
   it is possible to 
  design spin-transport devices based on graphene nanoribbons by introducing divacancies close to its edges by means of electron or ion irradiation.

\begin{acknowledgments} 

This work was supported by the Polish National Science Center (Grant  DEC-2011/03/B/ST3/00091), the Basque Government through the NANOMATERIALS project (Grant  IE05-151) under the 
ETORTEK Program {\it iNanogune}, the Spanish Ministry of Economy and Competitiveness MINECO (Grants FIS2013-48286-C2-1-P and FIS2012-33521), and the University of the Basque Country (Grant No. IT-366-07). WJ and LC acknowledge the DIPC for its
generous hospitality.

\end{acknowledgments}

\end{document}